# An Introduction to the NMPC-Graph as General Schema for Causal Modeling of Nonlinear, Multivariate, Dynamic, and Recursive Systems with Focus on Time-Series Prediction

Christoph Jahnz
ORCID: 0000-0003-3297-7564
CYNERELO Forecasting, Econometric Studies
Frankfurt am Main, Germany
christoph.jahnz@cynerelo.com

*Abstract*—**While the disciplines of physics and engineering sciences in many cases have taken advantage from accurate time-series prediction by applying ordinary differential equation systems upon precise basic physical laws such approach hardly could be adopted by other scientific disciplines where precise mathematical basic laws are unknown. A new modeling schema, the NMPC-graph, opens the possibility of interdisciplinary and generic nonlinear, multivariate, dynamic, and recursive causal modeling in domains where basic laws are only known as qualitative relationships among parameters while their precise mathematical nature remains undisclosed at modeling time. The symbolism of NMPC-graph is kept simple and suited for analysts without advanced mathematical skills.**

**This article presents the definition of the NMPC-graph modeling method and its six component types. Further, it shows how to solve the inverse problem of deriving a nonlinear ordinary differential equation system from any NMPC-graph in conjunction with historic calibration data by means of machine learning. This article further discusses how such a derived NMPC-model can be used for hypothesis testing and time-series prediction with the expectation of gaining prediction accuracy in comparison to conventional prediction methods.**

## I. Motivation for Introducing a New Modeling Method

Today's commonly established approaches of calculating the future behavior of complex systems whose parameters are expressed numerically roughly might be divided into four major categories of popular mathematical methods: First of all, there is the family of stochastic methods, secondly fundamental equation based simulations, thirdly methods of structural modeling and finally conventional inductive machine learning methods. Each of these method categories is characterized by specific and significant drawbacks concerning prediction accuracy of individual system behavior.

Stochastic models based on a collection of observations in the past resulting in probability based predictions [1] are widely spread. They usually abstract from system's internal causality. Thus, they are lacking information for predicting precise and non-probabilistic individual system behavior in phase space.

Simulations of continuous systems based on fundamental equations [2] are at the other end of system modeling. They do not abstract system's causality but specify it to a high level of detail given by mathematic equations including the specification of all mathematic constants. In cases where system's nature is not completely understood, simulation models tend to be over-specific and therefore likely to be inaccurate. This especially is the case for nonlinear systems where finding adequate equations and their constants is a sophisticated and costly undertaking.

Structural modeling here refers to the methods of *structural equation modeling* (SEM) [3] and *dynamic causal modeling* (DCM) [6] which both allow causal modeling without creating over-specific models. However, in both cases much of the theory is bound to linearity and where not, nonlinear relationships are quite predetermined and far away from being flexible [4][5]. There are more methods in the category of reasoning under uncertainty like *bayesian networks* [13] and *fuzzy logic* [14] which also could be classified as structural modeling methods. Although they disclose more of a system's internal nature at the root they tend to abstract system structure and rather deal with uncertainty instead of reflecting system's precise phase space characteristics.

Conventional supervised machine learning methods like *back-propagation* algorithm for *artificial neural networks* [7][10], training algorithms for *decision trees* [8][11], and *support vector machines* [9][12] rely on inductive classifying, i.e. most of the information they gain about the system to be represented is given by calibration data. Finding a way of how generalization has to take place is left to the specific algorithm. There hardly exist methodical standards for a human analyst to assist the algorithm in how these calibration data has to be interpreted. Thus, there needs to be a considerably large amount of calibration data and effort to force the algorithm into a lifelike generalization, otherwise the model is likely to become inaccurate.

In summary, all of the above methods are showing weakness in either including relevant knowledge from the analyst or reflecting sufficiently the system's structural essentials or coping with nonlinearity or even all of it.

One good example where the drawbacks of conventional methods become obvious is modeling in the discipline of macroeconomy. There, relevant time-series sets often contain only some dozens or hundreds of calibration records of historic data [15]. Taking into account the many economic parameters which play a role within one scenario the available amount of calibration records is likely to not be enough for representatively delivering evidence for all kinds of combined parameter influences. In other words, there is too few information within historic data for disclosing system's nature.

An accurate model for individual predictions would need to include additional knowledge. However, while analysts are aware of macroeconomic laws for which exist theoretical equations most of such equations are rather expected to only roughly approximate an aspect of some causality instead of modeling it precisely [16]. Furthermore, macroeconomic causality often implies nonlinearity which is not being strongly supported by many of the conventional modeling methods. So it is hard for contemporary analysts to contribute their macroeconomic knowledge into any conventional prediction model [17].

The new method introduced within this article allows analysts to specify their belief about causality by the symbolism of a new modeling schema which is the so called ***n**onlinear **m**onotone **p**arameter **c**oupling graph*, or shortly, the *NMPC-graph*.

A NMPC-graph expresses a network of causal relationships, potentially nonlinear, which explains parameters representing real numbers by other parameters and in case of recursivity, also by themselves. The symbolism is simple, non-algebraic, and universal. However, it comprises only six basic component types which are defined in section II. The characteristics of a well-formed NMPC-graph are presented in section IV.A.

A nonlinear relationship is rather expressed qualitatively, e.g. one component might express "the faster first input value grows and the faster the second input value decreases the faster the output grows" while the other might express "the higher the input value the faster the output value decreases". Together, such relationships form an abstract ordinary differential equation system consisting of strictly monotone univariate and/or multivariate functions which might be combined with integral and differential functions sharing one independent parameter which typically represents time. A NMPC-graph example is shown in section III.B. The precise shape of all monotone functions remains undefined within any NMPC-graph. Their shape is only known within a NMPC-model after calibration process as described in section 0.

Actually, the design of the NMPC-graph symbolism was kept in such way that also analysts who are not familiar with advanced mathematics or system theory are provided with an easy-to-use method for sophisticated system modeling. Conversely, any software converting NMPC-graphs into NMPC-models and evaluating them will carry a high mathematical load. One of the principle motivations for introducing the NMPC-graph method was role-sharing in modeling work, meaning each one, human analysts and computers, do what they can do best, respectively.

A specific NMPC-graph is a hypothesis about some system. Together with data representing system's history it can be converted into a NMPC-model as shown in section 0. Hypothesis testing is discussed in section V and time-series prediction in section VI.

A glossary with important terms of NMPC methodology and mathematic variables can be found in the appendix.

## II. Component Definitions

NMPC-graphs can be composed by up to six basic component types which are described in following.

### A. Parameter

The first component type is the *NMPC-parameter*, or shortly, *parameter*. Each parameter expresses a scalar real number dependent on the independent parameter $t$ and further on will be referred to by variable $p$. For simplicity, due to its most common application, $t$ in following will be named *time parameter* although it might also refer to an independent quantity with other meaning than time.

Within a NMPC-graph parameters describe the system's *observable* or *latent* state at one point of $t$, i.e. typically at one point in time.

$$t \in \mathbb{R} \; ; p(t) \in \mathbb{R} \; ; p_{min} \leq p(t) \leq p_{max} \qquad (1)$$

Within a NMPC-graph each *parameter* can serve as input of one or multiple NMPC-components or it can serve as output of one other NMPC-component or both, simultaneously.

<u>Symbolism:</u>

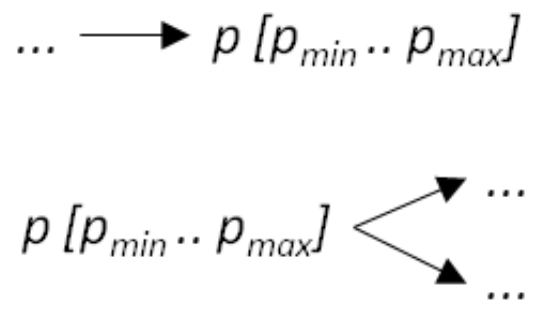

### B. Bilateral couplings

There exist three types of so called *bilateral couplings*. A bilateral coupling expresses an optionally temporal and basically nonlinear relationship between one input $x_{in}$ and one output $x_{out}$. Input and output represent scalar real number values. Within a NMPC-graph, input and output of each bilateral coupling has to be connected to parameters or to components of any other NMPC-component type.

The nonlinear relationship of all types of bilateral coupling is defined in two fundamental aspects: (a) Bilateral couplings describe always the positive or negative growth of output $x_{out}$ over time. (b) The nonlinear relationship between input $x_{in}$ and output $x_{out}$ is always abstracted by a so called *m-function* whose precise shape remains unknown at NMPC-graph design time. The *m-function* $m(x)$ is strictly monotone. $t$ is the independent parameter.

For all bilateral couplings applies following:

$$t \in \mathrm{IR}\ ;\ x_{in}(t) \in \mathrm{IR}\ ;\ x_{in_min} \le x_{in}(t) \le x_{in_max}\ ;\ x_{out} \in \mathrm{IR}\ ;$$

$$x_{out_min} \le x_{out}(t) \le x_{out_max}$$

$$x \in \mathrm{IR}\ ;\ m(x) \in \mathrm{IR}\ ;\ \frac{\partial m(x)}{\partial x} > 0\ . \tag{2}$$

Generally, bilateral couplings are defined verbally and additionally by a mathematic representation which is able to represent the verbal definition in a formal way. This formal description, however, is not necessarily entirely represented by the verbal definition.

*1) Integrative coupling*

The *m-function* for *integrative couplings* is termed $\tilde{i}(x)$ with

$$x \in \mathrm{IR}\ ;\ \tilde{i}(x) \in \mathrm{IR}\ ;\ \frac{\partial \tilde{i}(x)}{\partial x} > 0\ . \tag{3}$$

Three coupling variants have to be distinguished:

*a)* The larger $x_{in}$, the faster $x_{out}$ grows / the slower $x_{out}$ decreases. The smaller $x_{in}$, the faster $x_{out}$ decreases / the slower $x_{out}$ grows.

Symbolism:

$x_{in}\ [x_{in_min} .. x_{in_max}]$ ↑↗ ⟶ $x_{out}\ [x_{out_min} .. x_{out_max}]$

Equation:

$$\frac{\partial x_{out}(t)}{\partial t} = \tilde{i}(x_{in}(t))\ ;$$

$$x_{out}(t) = \int \tilde{i}(x_{in}(t))\partial t = i_{++}(x_{in}(t)) \tag{4}$$

*b)* The larger $x_{in}$, the faster $x_{out}$ decreases / the slower $x_{out}$ grows. The smaller $x_{in}$, the faster $x_{out}$ grows / the slower $x_{out}$ decreases.

Symbolism:

$x_{in}\ [x_{in_min} .. x_{in_max}]$ ↑↘ ⟶ $x_{out}\ [x_{out_min} .. x_{out_max}]$

Equation:

$$\frac{\partial x_{out}(t)}{\partial t} = -\tilde{i}(x_{in}(t))\ ;$$

$$x_{out}(t) = -\int \tilde{i}(x_{in}(t))\partial t = i_{+-}(x_{in}(t)) \tag{5}$$

*c)* The smaller $x_{in}$, the faster $x_{out}$ grows / the slower $x_{out}$ decreases. The larger $x_{in}$, the faster $x_{out}$ decreases / the slower $x_{out}$ grows.

Symbolism:

$x_{in}\ [x_{in_min} .. x_{in_max}]$ ↓↗ ⟶ $x_{out}\ [x_{out_min} .. x_{out_max}]$

Equation:

$$\frac{\partial x_{out}(t)}{\partial t} = \tilde{i}(-x_{in}(t))\ ;$$

$$x_{out}(t) = \int \tilde{i}(-x_{in}(t))\partial t = i_{-+}(x_{in}(t)) \tag{6}$$

In following *integrative couplings* will be referred to by a generalized function *i(...)* defined as

$$i(...) \supseteq i_{++}(...);\ i(...) \supseteq i_{+-}(...);\ i(...) \supseteq i_{-+}(...)\ .$$

*2) Synchronous coupling*

The *m-function* for *synchronous couplings* is termed $\tilde{s}(x)$ with

$$x \in \mathrm{IR}\ ;\ \tilde{s}(x) \in \mathrm{IR}\ ;\ \frac{\partial \tilde{s}(x)}{\partial x} > 0\ . \tag{7}$$

Three coupling variants have to be distinguished:

*a)* The faster $x_{in}$ grows, the faster $x_{out}$ grows / the slower $x_{out}$ decreases. The faster $x_{in}$ decreases, the faster $x_{out}$ decreases / the slower $x_{out}$ grows.

Symbolism:

$x_{in}\ [x_{in_min} .. x_{in_max}]$ ↗↗ ⟶ $x_{out}\ [x_{out_min} .. x_{out_max}]$

Equation:

$$\frac{\partial x_{out}(t)}{\partial t} = \frac{\partial \tilde{s}(x_{in}(t))}{\partial t}$$

$$x_{out}(t) = \tilde{s}(x_{in}(t)) = s_{++}(x_{in}(t)) \tag{8}$$

*b)* The faster $x_{in}$ grows, the faster $x_{out}$ decreases. The faster $x_{in}$ decreases, the faster $x_{out}$ grows / the slower $x_{out}$ decreases.

Symbolism:

$x_{in}\ [x_{in_min} .. x_{in_max}]$ ↗↘ ⟶ $x_{out}\ [x_{out_min} .. x_{out_max}]$

Equation:

$$\frac{\partial x_{out}(t)}{\partial t} = -\frac{\partial \tilde{s}(x_{in}(t))}{\partial t}$$

$$x_{out}(t) = -\tilde{s}(x_{in}(t)) = s_{+-}(x_{in}(t)) \tag{9}$$

*c)* The faster $x_{in}$ decreases, the faster $x_{out}$ grows / the slower $x_{out}$ decreases. The faster $x_{in}$ grows, the faster $x_{out}$ decreases / the slower $x_{out}$ grows.

Symbolism:

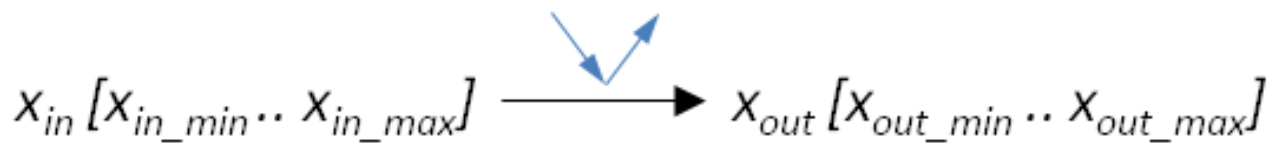

Equation:

$$\frac{\partial x_{out}(t)}{\partial t} = -\frac{\partial \tilde{s}(x_{in}(t))}{\partial t} \tag{10}$$

$$x_{out}(t) = -\tilde{s}(x_{in}(t)) = s_{-+}(x_{in}(t))$$

In following *synchronous couplings* will be referred to by a generalized function *s(...)* defined as

$$s(...) \supseteq s_{++}(...);\ s(...) \supseteq s_{+-}(...)\ ;\ s(...) \supseteq s_{-+}(...)\ .$$

*3) Differential coupling*

The *m-function* for *differential couplings* is termed $\tilde{d}(x)$ with

$$x \in \text{IR}\ ;\ \tilde{d}(x) \in \text{IR};\ \frac{\partial \tilde{d}(x)}{\partial x} > 0\ . \tag{11}$$

Three coupling variants have to be distinguished:

*a)* The more $x_{in}$ accelerates, the faster $x_{out}$ grows / the slower $x_{out}$ decreases. The more $x_{in}$ decelerates, the faster $x_{out}$ decreases / the slower $x_{out}$ grows.

Symbolism:

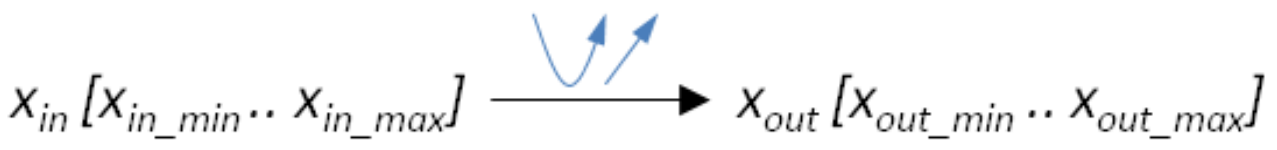

Equation:

$$\frac{\partial x_{out}(t)}{\partial t} = \frac{\partial \tilde{d}\left(\frac{\partial x_{in}(t)}{\partial t}\right)}{\partial t}$$

$$x_{out}(t) = \tilde{d}\left(\frac{\partial x_{in}(t)}{\partial t}\right) = d_{++}(x_{in}(t)) \tag{12}$$

*b)* The more $x_{in}$ accelerates, the faster $x_{out}$ decreases / the slower $x_{out}$ grows. The more $x_{in}$ decelerates, the faster $x_{out}$ grows / the slower $x_{out}$ decreases.

Symbolism:

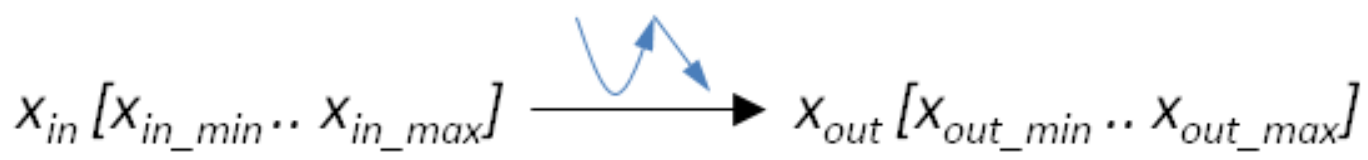

Equation:

$$\frac{\partial x_{out}(t)}{\partial t} = -\frac{\partial \tilde{d}\left(\frac{\partial x_{in}(t)}{\partial t}\right)}{\partial t}$$

$$x_{out}(t) = -\tilde{d}\left(\frac{\partial x_{in}(t)}{\partial t}\right) = d_{+-}(x_{in}(t)) \tag{13}$$

*c)* The more $x_{in}$ decelerates, the faster $x_{out}$ grows / the slower $x_{out}$ decreases. The more $x_{in}$ accelerates, the faster $x_{out}$ decreases / the slower $x_{out}$ grows.

Symbolism:

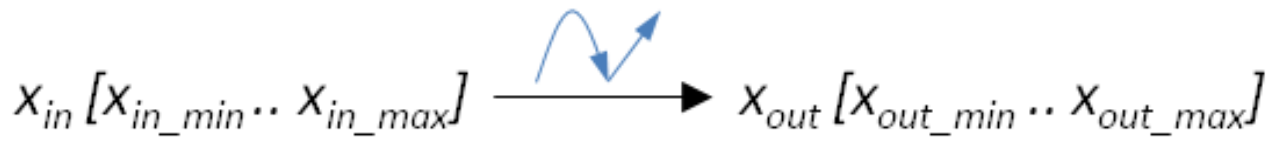

Equation:

$$\frac{\partial x_{out}(t)}{\partial t} = -\frac{\partial \tilde{d}\left(\frac{\partial x_{in}(t)}{\partial t}\right)}{\partial t} \tag{14}$$

$$x_{out}(t) = -\tilde{d}\left(\frac{\partial x_{in}(t)}{\partial t}\right) = d_{-+}(x_{in}(t))$$

In following *differential couplings* will be referred to by a generalized function *d(...)* defined as

$$d(...) \supseteq d_{++}(...);\ d(...) \supseteq d_{+-}(...)\ ;\ d(...) \supseteq d_{-+}(...)\ .$$

## C. Operators

In cases where exist more than one input affecting a NMPC-component, *operators* are needed. A NMPC-graph allows two types of operators, *summators* and *modulators*. An operator is a NMPC-component.

*1) Summator*

A *summator* reflects the linear arithmetic sum of any number of signed inputs from other NMPC-component outputs returned by one output. This way, it effectively calculates sums and differences. The sign of each input needs to be specified by an inclined (if input adds positive) or declined (if input adds negative) *indicator arrow*. Each input and the output of a summator must be connected to a NMPC-component.

Summators do not involve nonlinear transformation like implied by the *m-function* of bilateral couplings.

Summators are eligible in cases where there have to be considered multiple influences represented by *j* state variables $x_{in,v}(t)$ affecting one target state variable $x_{out}(t)$ independently from each other.

$$x_{in,1}(t),\ x_{in,2}(t)\ \ldots\ ,\ x_{in,j}(t) \in \text{IR}\ ;$$

$$x_{in,v_min} \le x_{in,v}(t) \le x_{in,v_max}\ ;$$

$$x_{out} \in \text{IR}\ ;\ x_{out_min} \le x_{out}(t) \le x_{out_max}$$

Symbolism:

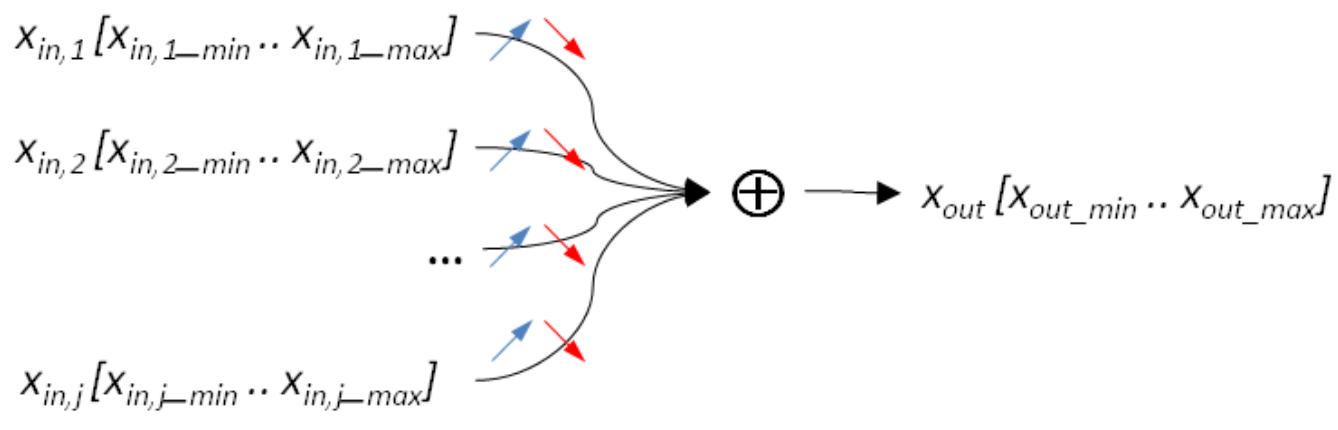

Equation:

$$x_{out}(t) = \sum_{v=1}^{j} \pm x_{in,v}(t)$$
$$= \sum_{v=1}^{j} \tilde{a}_v(x_{in,v}(t))$$
$$= a(\tilde{a}_1(x_{in,1}(t)),\ \tilde{a}_2(x_{in,2}(t)),\ \dots,\ \tilde{a}_j(x_{in,j}(t))) \qquad (15)$$

The function $\tilde{a}_v(x)$ reflects the sign of the summator's input $v$. The function $a(...) \in \mathrm{IR}$ represents the summator as a whole with $j$ input state variables.

*2) Modulator*

A modulator expresses an output $x_{out}$ depending on both one input $x_{in1}$ and one input $x_{in2}$, simultaneously. Each input must be associated to one *indicator arrow*. An indicator arrow pointing upwards inclined means that a rising value of respective input will cause the output value to also rise. An indicator arrow pointing downwards declined means that a rising value of respective input will cause the output value to decrease.

Presented in a Cartesian coordinate plane a modulator is a continuous function of two variables where the sign of surface inclination always remains unchanged when moving over the surface in parallel to one of its input axes. The shape of the surface can be linear or nonlinear and actually be of any form as long as continuity and the inclination constraints are fulfilled.

When changing only one input value while leaving the other one constant the modulator behaves like a *synchronous coupling* (see section II.B.2) concerning the changing input and the output.

In contrary to summators which are used for expressing various inputs influencing the output independently from each other modulators are used to express one input influencing the output while being conditioned by another input.

$$x_{in,1}(t),\ x_{in,2}(t) \in \mathrm{IR}\ ;$$
$$x_{in1_min} \le x_{in1}(t) \le x_{in1_max}\ ;$$
$$x_{in2_min} \le x_{in2}(t) \le x_{in2_max}\ ;$$
$$x_{out} \in \mathrm{IR}\ ;\ x_{out_min} \le x_{out}(t) \le x_{out_max}$$

Three modulator variants have to be distinguished:

*a)* The faster $x_{in1}$ grows, the faster $x_{out}$ grows. The faster $x_{in2}$ grows, the faster $x_{out}$ grows. $x_{in1}$ conditions the influence of $x_{in2}$ and vice versa.

Symbolism:

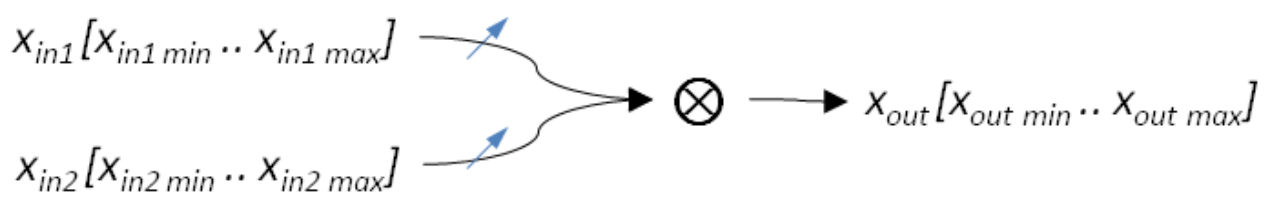

Equation:

$$x_{out}(t) = m_{++}(x_{in1}(t), x_{in2}(t));$$
$$\frac{\partial m_{++}(x_{in1}, x_{in2})}{\partial x_{in1}} > 0;\ \frac{\partial m_{++}(x_{in1}, x_{in2})}{\partial x_{in2}} > 0 \qquad (16)$$

*b)* The faster $x_{in1}$ grows, the faster $x_{out}$ grows. The faster $x_{in2}$ decreases, the faster $x_{out}$ grows. $x_{in1}$ conditions the influence of $x_{in2}$ and vice versa.

Symbolism:

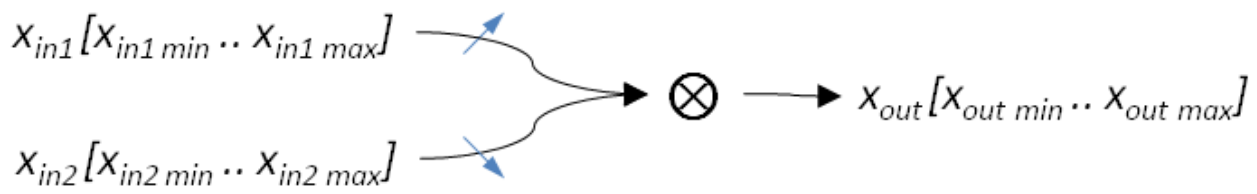

Equation:

$$x_{out}(t) = m_{+-}(x_{in1}(t), x_{in2}(t));$$
$$\frac{\partial m_{+-}(x_{in1}, x_{in2})}{\partial x_{in1}} > 0;\ \frac{\partial m_{+-}(x_{in1}, x_{in2})}{\partial x_{in2}} < 0 \qquad (17)$$

*c)* The faster $x_{in1}$ decreases, the faster $x_{out}$ grows. The faster $x_{in2}$ decreases, the faster $x_{out}$ grows. $x_{in1}$ conditions the influence of $x_{in2}$ and vice versa.

Symbolism:

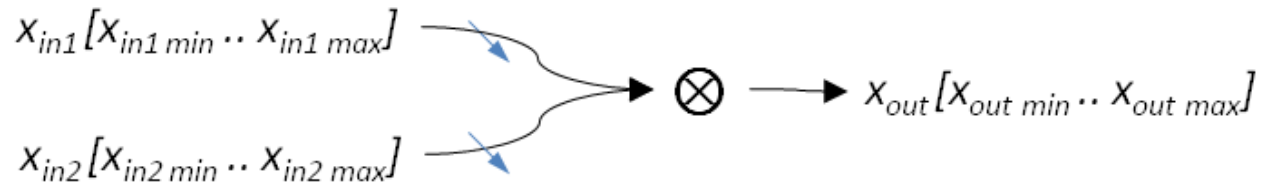

Equation:

$$x_{out}(t) = m_{--}(x_{in1}(t), x_{in2}(t));$$
$$\frac{\partial m_{--}(x_{in1}, x_{in2})}{\partial x_{in1}} < 0;\ \frac{\partial m_{--}(x_{in1}, x_{in2})}{\partial x_{in2}} < 0 \qquad (18)$$

In following *modulators* will be referred to by a generalized function $m(...)$ defined as

$$m(...) \supseteq m_{++}(...);\ m(...) \supseteq m_{+-}(...)\ ;\ m(...) \supseteq m_{--}(...)\quad .$$

## III. Nodal Matrix Definition of the NMPC-Graph

### *A. Definition of the state equation*

Given a set of nodes where each input and each output of any component of a given NMPC-graph is connected to one of these nodes, respectively, and each node represents a state variable $x_v(t)$ the resulting equation system can be expressed in matrix notation as differential equation system by referring to the definitions in section II as

$$\frac{\partial}{\partial t}\begin{bmatrix} x_1(t) \\ x_2(t) \\ \vdots \\ x_n(t) \end{bmatrix} = \begin{bmatrix} \tilde{i}_{1,1}(x_1) & \tilde{i}_{1,2}(x_2) & \cdots & \tilde{i}_{1,n}(x_n) \\ \tilde{i}_{2,1}(x_1) & \tilde{i}_{2,2}(x_2) & \cdots & \tilde{i}_{2,n}(x_n) \\ \vdots & \vdots & \ddots & \vdots \\ \tilde{i}_{n,1}(x_1) & \tilde{i}_{n,2}(x_2) & \cdots & \tilde{i}_{n,n}(x_n) \end{bmatrix} \cdot \begin{bmatrix} 1 \\ 1 \\ \vdots \\ 1 \end{bmatrix}$$

$$+\frac{\partial}{\partial t}\begin{bmatrix} \tilde{s}_{1,1}(x_1) & \tilde{s}_{1,2}(x_2) & \cdots & \tilde{s}_{1,n}(x_n) \\ \tilde{s}_{2,1}(x_1) & \tilde{s}_{2,2}(x_2) & \cdots & \tilde{s}_{2,n}(x_n) \\ \vdots & \vdots & \ddots & \vdots \\ \tilde{s}_{n,1}(x_1) & \tilde{s}_{n,2}(x_2) & \cdots & \tilde{s}_{n,n}(x_n) \end{bmatrix} \cdot \begin{bmatrix} 1 \\ 1 \\ \vdots \\ 1 \end{bmatrix}$$

$$+\frac{\partial}{\partial t}\begin{bmatrix} \tilde{d}_{1,1}(\dot{x}_1) & \tilde{d}_{1,2}(\dot{x}_2) & \cdots & \tilde{d}_{1,n}(\dot{x}_n) \\ \tilde{d}_{2,1}(\dot{x}_1) & \tilde{d}_{2,2}(\dot{x}_2) & \cdots & \tilde{d}_{2,n}(\dot{x}_n) \\ \vdots & \vdots & \ddots & \vdots \\ \tilde{d}_{n,1}(\dot{x}_1) & \tilde{d}_{n,2}(\dot{x}_2) & \cdots & \tilde{d}_{n,n}(\dot{x}_n) \end{bmatrix} \cdot \begin{bmatrix} 1 \\ 1 \\ \vdots \\ 1 \end{bmatrix}$$

$$+\frac{\partial}{\partial t}\begin{bmatrix} \tilde{a}_{1,1}(x_1) & \tilde{a}_{1,2}(x_2) & \cdots & \tilde{a}_{1,n}(x_n) \\ \tilde{a}_{2,1}(x_1) & \tilde{a}_{2,2}(x_2) & \cdots & \tilde{a}_{2,n}(x_n) \\ \vdots & \vdots & \ddots & \vdots \\ \tilde{a}_{n,1}(x_1) & \tilde{a}_{n,2}(x_2) & \cdots & \tilde{a}_{n,n}(x_n) \end{bmatrix} \cdot \begin{bmatrix} 1 \\ 1 \\ \vdots \\ 1 \end{bmatrix}$$

$$+\frac{\partial}{\partial t}\begin{bmatrix} m_{1,1}(x_1,x_2) & m_{1,2}(x_1,x_3) & \cdots & m_{1,n}(x_1,x_n) & m_{1,n+1}(x_2,x_3) & m_{1,n+2}(x_2,x_4) & \cdots & m_{1,q}(x_{n-1},x_n) \\ m_{2,1}(x_1,x_2) & m_{2,2}(x_1,x_3) & \cdots & m_{2,n}(x_1,x_n) & m_{2,n+1}(x_2,x_3) & m_{2,n+2}(x_2,x_4) & \cdots & m_{2,q}(x_{n-1},x_n) \\ \vdots & \vdots & & \vdots & \vdots & \vdots & \ddots & \vdots \\ m_{n,1}(x_1,x_2) & m_{n,2}(x_1,x_3) & \cdots & m_{n,n}(x_1,x_n) & m_{n,n+1}(x_2,x_3) & m_{n,n+2}(x_2,x_4) & \cdots & m_{n,q}(x_{n-1},x_n) \end{bmatrix} \cdot \begin{bmatrix} 1 \\ 1 \\ \vdots \\ 1 \\ 1 \\ 1 \\ \vdots \\ 1 \end{bmatrix} \quad (19)$$

which per definition is equivalent to

$$X = I \cdot \begin{bmatrix} 1 \\ 1 \\ \vdots \\ 1 \end{bmatrix} + S \cdot \begin{bmatrix} 1 \\ 1 \\ \vdots \\ 1 \end{bmatrix} + D \cdot \begin{bmatrix} 1 \\ 1 \\ \vdots \\ 1 \end{bmatrix} + A + M \cdot \begin{bmatrix} 1 \\ 1 \\ 1 \\ 1 \\ 1 \\ \vdots \\ 1 \end{bmatrix} \quad (20)$$

where $X$ represents a column vector with $n$ elements, each of them representing a NMPC-component's output value at one point in time, $I$ representing a $n \times n$ matrix for all *integrative couplings*, $S$ for all *synchronous couplings*, $A$ for *summators*, $D$ for all *differential couplings*, and, as exception, $M$ is the $n \times q$ matrix with $q = \begin{pmatrix} n \\ k \end{pmatrix}$ (*n choose k,* i.e. each of all possible combinations of two input state variables into one *modulator*, respectively) with $k$=2 for all *modulators* within the NMPC-graph. Each of those matrices is multiplied by a column vector consisting of ones allowing the matrix notation being converted into an algebraic equation system.

*NMPC-parameters* are a subset of $X$ being expressed by column vector $P$ where each parameter holds one number value for each point in time:

$$X \supseteq P \text{ with } x_1(t), x_2(t), \ldots, x_n(t)$$
$$\text{and } p_1(t), p_2(t), \ldots, p_z(t) \in \mathbb{R} \quad (21)$$

Referring to the definitions in section II and (20) with $t$ as *independent variable* and with all *state variables* $x_v(t)$ the matrix notation can be also written as

$$
\begin{bmatrix} x_1(t) \\ x_2(t) \\ \vdots \\ x_n(t) \end{bmatrix} =
\begin{bmatrix} i_{1,1}(x_1) & i_{1,2}(x_2) & \cdots & i_{1,n}(x_n) \\ i_{2,1}(x_1) & i_{2,2}(x_2) & \cdots & i_{2,n}(x_n) \\ \vdots & \vdots & \ddots & \vdots \\ i_{n,1}(x_1) & i_{n,2}(x_2) & \cdots & i_{n,n}(x_n) \end{bmatrix} \cdot \begin{bmatrix} 1 \\ 1 \\ \vdots \\ 1 \end{bmatrix}
+ \begin{bmatrix} s_{1,1}(x_1) & s_{1,2}(x_2) & \cdots & s_{1,n}(x_n) \\ s_{2,1}(x_1) & s_{2,2}(x_2) & \cdots & s_{2,n}(x_n) \\ \vdots & \vdots & \ddots & \vdots \\ s_{n,1}(x_1) & s_{n,2}(x_2) & \cdots & s_{n,n}(x_n) \end{bmatrix} \cdot \begin{bmatrix} 1 \\ 1 \\ \vdots \\ 1 \end{bmatrix}
+ \begin{bmatrix} d_{1,1}(x_1) & d_{1,2}(x_2) & \cdots & d_{1,n}(x_n) \\ d_{2,1}(x_1) & d_{2,2}(x_2) & \cdots & d_{2,n}(x_n) \\ \vdots & \vdots & \ddots & \vdots \\ d_{n,1}(x_1) & d_{n,2}(x_2) & \cdots & d_{n,n}(x_n) \end{bmatrix} \cdot \begin{bmatrix} 1 \\ 1 \\ \vdots \\ 1 \end{bmatrix}
+ \begin{bmatrix} a_1(\tilde{a}_{1,1}(x_1), \tilde{a}_{1,2}(x_2), \dots, \tilde{a}_{1,n}(x_n)) \\ a_2(\tilde{a}_{2,1}(x_1), \tilde{a}_{2,2}(x_2), \dots, \tilde{a}_{2,n}(x_n)) \\ \vdots \\ a_n(\tilde{a}_{n,1}(x_1), \tilde{a}_{n,2}(x_2), \dots, \tilde{a}_{n,n}(x_n)) \end{bmatrix}
+ \begin{bmatrix} m_{1,1}(x_1,x_2) & m_{1,2}(x_1,x_3) & \cdots & m_{1,n}(x_1,x_n) & m_{1,n+1}(x_2,x_3) & m_{1,n+2}(x_2,x_4) & \cdots & m_{1,q}(x_{n-1},x_n) \\ m_{2,1}(x_1,x_2) & m_{2,2}(x_1,x_3) & \cdots & m_{2,n}(x_1,x_n) & m_{2,n+1}(x_2,x_3) & m_{2,n+2}(x_2,x_4) & \cdots & m_{2,q}(x_{n-1},x_n) \\ \vdots & \vdots & & \vdots & \vdots & \vdots & \ddots & \vdots \\ m_{n,1}(x_1,x_2) & m_{n,2}(x_1,x_3) & \cdots & m_{n,n}(x_1,x_n) & m_{n,n+1}(x_2,x_3) & m_{n,n+2}(x_2,x_4) & \cdots & m_{n,q}(x_{n-1},x_n) \end{bmatrix} \cdot \begin{bmatrix} 1 \\ 1 \\ \vdots \\ 1 \\ 1 \\ 1 \\ \vdots \\ 1 \end{bmatrix} \quad (22)
$$

where each of the elements $i_{v,w}$, $s_{v,w}$, $d_{v,w}$, $a_v$, and $m_{v,w}$ with $v$, $w$ as matrix index represent either a *bilateral coupling* or an *operator* or the value 0, latter indicating a missing NMPC-component at this place.

*Modulators* are represented by $m_{v,w}$, enumerating all possible input variable combinations without repetitions. Direct feedbacks of a modulator's output into one of its inputs is not allowed, meaning that within a matrix row $r$ all elements

$m_{r,u}(x_v, x_w) = 0$ where $v = r$ or $w = r$.

Accordingly, direct feedback of a *synchronous coupling's* output into its input is not allowed, meaning that within a matrix row $r$ all elements $s_{r,u}(x_r) = 0$.

*Summators* are represented by one sum of state variables as function $a_v(\tilde{a}_{v,1}(x_1), \tilde{a}_{v,2}(x_2), \dots, \tilde{a}_{v,n}(x_n))$ per state variable $x_v$. Direct feedback from a summators output into its input is not allowed, meaning that within a matrix row $r$ the summator's input $\tilde{a}_{r,w}(x_r) = 0$.

Further, each of all $n$ state variables in $X$ represents the output node of at most one NMPC-function:

$$\left.\begin{array}{llllll}\forall i \,\big|_{i_{v,w}\neq 0} & \rightarrow\; i_{v,u}=0, & s_{v,j}=0, & d_{v,j}=0, & a_v=0, & m_{v,l}=0 \\ \forall s \,\big|_{s_{v,w}\neq 0} & \rightarrow\; i_{v,j}=0, & s_{v,u}=0, & d_{v,j}=0, & a_v=0, & m_{v,l}=0 \\ \forall d \,\big|_{d_{v,w}\neq 0} & \rightarrow\; i_{v,j}=0, & s_{v,j}=0, & d_{v,u}=0, & a_v=0, & m_{v,l}=0 \\ \forall a \,\big|_{a_v\neq 0} & \rightarrow\; i_{v,j}=0, & s_{v,j}=0, & d_{v,j}=0, & & m_{v,l}=0 \\ \forall m \,\big|_{m_{v,w}\neq 0} & \rightarrow\; i_{v,j}=0, & s_{v,j}=0, & d_{v,j}=0, & a_v=0, & m_{v,u}=0 \end{array}\right|\; u\neq w;\, j=1..n;\, l=1..q \qquad (23)$$

*B. Example of a NMPC-Graph*

Following example illustrates how a NMPC-graph is represented in matrix notation. Here a hypothetic scenario of a time dependent measure of soil moisture is shown which depends on natural precipitation and evaporation:

The higher the soil moisture ($p_3$) the more intensive the evaporation ($x_4$), amplified ($m_{4,2}$) with growing sunshine intensity ($p_1$). Evaporation is subtracted ($a_6$) from a unit of measure adjusted precipitation ($x_5$). The unit of measure adjustment ($S_{5,2}$) is applied to original precipitation ($p_2$). The more positive this result, the faster soil moisture grows ($i_{3,6}$). The more negative this result, the faster soil moisture decreases ($i_{3,6}$).

By convention, all measurable parameters are shown underlined while parameters which are believed to be meaningful to the scenario but cannot directly be measured, i.e. latent parameters, are shown in parenthesis:

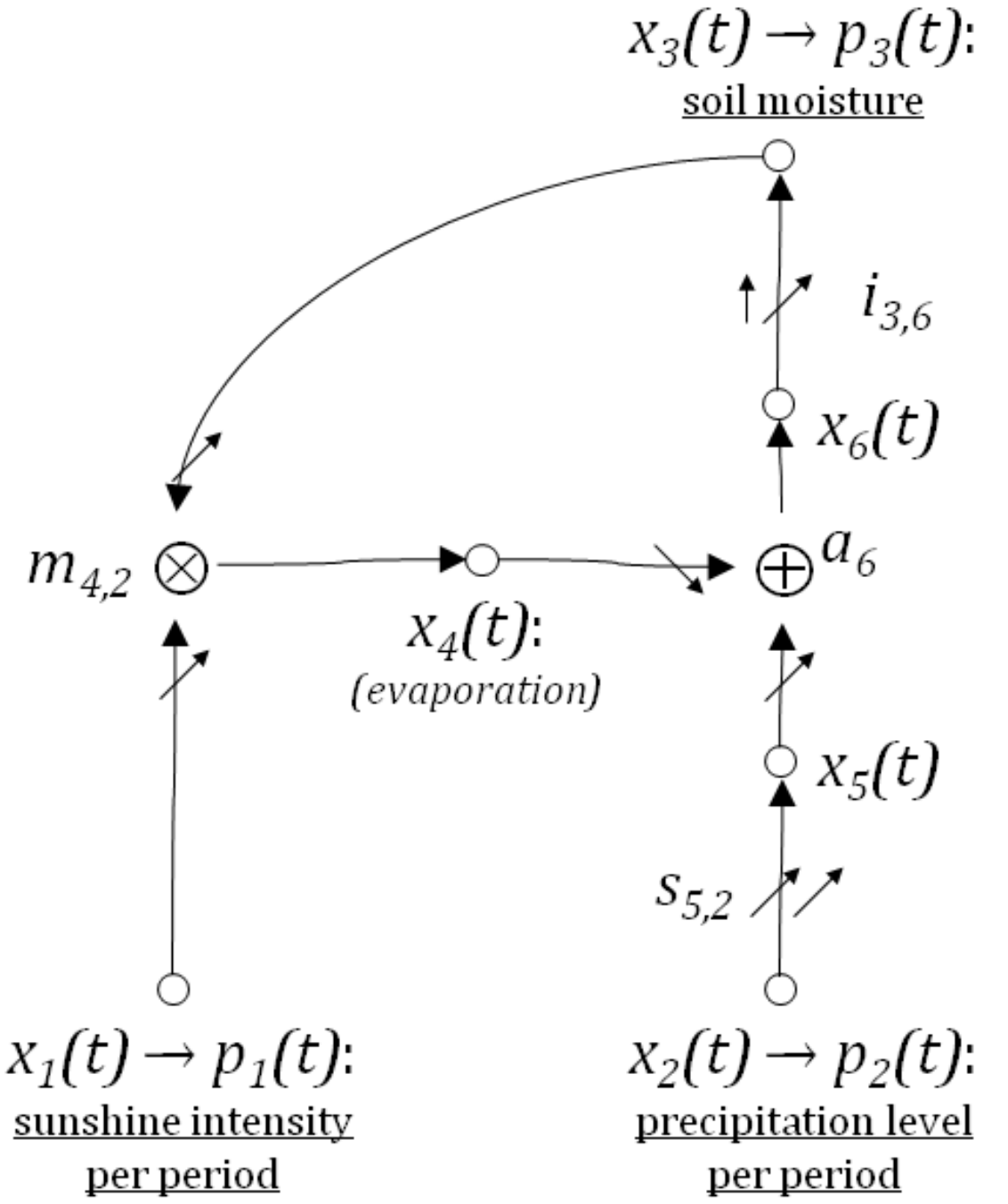

The corresponding matrix equation looks like following:

$$
\begin{bmatrix} x_1(t)=p_1(t) \\ x_2(t)=p_2(t) \\ x_3(t)=p_3(t) \\ x_4(t) \\ x_5(t) \\ x_6(t) \end{bmatrix}
=
\int \begin{bmatrix} 0&0&0&0&0&0 \\ 0&0&0&0&0&0 \\ 0&0&0&0&0&i_{3,6}(x_6(t)) \\ 0&0&0&0&0&0 \\ 0&0&0&0&0&0 \\ 0&0&0&0&0&0 \end{bmatrix} \partial t \cdot \begin{bmatrix} 1\\1\\1\\1\\1\\1 \end{bmatrix}
$$

$$
+ \begin{bmatrix} 0&0&0&0&0&0 \\ 0&0&0&0&0&0 \\ 0&0&0&0&0&0 \\ 0&0&0&0&0&0 \\ 0&s_{5,2}(x_2(t))&0&0&0&0 \\ 0&0&0&0&0&0 \end{bmatrix} \cdot \begin{bmatrix} 1\\1\\1\\1\\1\\1 \end{bmatrix}
$$

$$
+ \begin{bmatrix} 0&0&0&0&0&0 \\ 0&0&0&0&0&0 \\ 0&0&0&0&0&0 \\ 0&0&0&0&0&0 \\ 0&0&0&0&0&0 \\ 0&0&0&0&0&0 \end{bmatrix} \cdot \begin{bmatrix} 1\\1\\1\\1\\1\\1 \end{bmatrix}
$$

$$
+ \begin{bmatrix} 0&0&0&0&0&0 \\ 0&0&0&0&0&0 \\ 0&0&0&0&0&0 \\ 0&0&0&0&0&0 \\ 0&0&0&0&0&0 \\ 0&0&0&\tilde{a}_{6,4}(-x_4(t))&\tilde{a}_{6,5}(x_5(t))&0 \end{bmatrix} \cdot \begin{bmatrix} 1\\1\\1\\1\\1\\1 \end{bmatrix}
$$

$$
+ \begin{bmatrix} 0&0&0&0&0&0&0&0&0&0&0&0&0&0&0&0 \\ 0&0&0&0&0&0&0&0&0&0&0&0&0&0&0&0 \\ 0&0&0&0&0&0&0&0&0&0&0&0&0&0&0&0 \\ 0&m_{4,2}(x_1(t),x_3(t))&0&0&0&0&0&0&0&0&0&0&0&0&0&0 \\ 0&0&0&0&0&0&0&0&0&0&0&0&0&0&0&0 \\ 0&0&0&0&0&0&0&0&0&0&0&0&0&0&0&0 \end{bmatrix} \cdot \begin{bmatrix} 1\\1\\1\\1\\1\\1\\1\\1\\1\\1\\1\\1\\1\\1\\1 \end{bmatrix} \tag{24}
$$

with $p_1(t) = x_1(t)$, $p_2(t) = x_2(t)$, and $p_3(t) = x_3(t)$.

## IV. Converting the NMPC-Graph Into a NMPC-Model by Calibration with Historic Data

### A. The principle of calibration

Modeling dynamic systems by NMPC-graphs is suited for such systems where types of NMPC-couplings between parameters are believed to be known, but not the precise shapes of their monotone and possibly nonlinear functions.

As the precise shapes of monotone functions *i, s, d,* and *m* in sections II.B and II.C are not fully defined in their input variables, the differential state equation system given by (19), (20), and (22) is neither. For calculating the evolution of state variables *X* in (19) over independent parameter *t* precise manifestations need to be determined for all monotone functions within the NMPC-graph. This is known as an *inverse problem* [18][19] and can be achieved numerically by a procedure which in following will be referred to as *calibration*.

*Calibration* is a network-suited regression method based on an iterative nonlinear regression calculus with a *forward pass* and a *backward pass* for each cycle while applying modulated noise [22] in order to overcome local minima. The objective is to convert a specific NMPC-graph into a NMPC-model, which is a differential state equation system corresponding to (22) and fully defined in its state variables *X* and independent parameter *t*. Thus, a NMPC-model is fully defined by its NMPC-parameters, NMPC-functions with their regression constants, and their couplings.

The NMPC-functions which need to be determined by a regression method are all of the bilateral couplings and modulators while NMPC-parameters and summators are already fully defined without calibration.

Let $u$ be an index uniquely identifying each NMPC-component in an NMPC-graph. Then the time independent regression constants for all **bilateral couplings** can be determined by some regression method [20][21]

$$(k_{u,1}, k_{u,2}, \dots, k_{u,h(u)}) = regression_u(x_{in,u}(t_j), x_{out,u}(t_j))\Big|_{t_j=t_{min}\dots t_{max}} \quad (25)$$

with a number of $h(u)$ regression constants $k_u$ of a bilateral coupling $u$.

Since the regression constants define either of the monotone functions $\tilde{d}(x)$, $\tilde{i}(x)$, and $\tilde{s}(x)$ they can be calculated as long as for each calibratable NMPC-function there is a sufficient large number of calibration sets containing time dependent values for input and output state variables with a sufficient coverage of the relevant value space for input and output state variable. Calibration sets for bilateral couplings then are represented by

$$x_{in,u}(t_j)\ ;\ x_{out,u}(t_j)\ \Big|_{t_j=t_{min}\dots t_{max}} \quad (26)$$

Here, all $x_{in,u}(t_j)$ can be calculated in a *forward pass* by (29) and all $x_{out,u}(t_j)$ can be calculated within a *backward pass* by (30) (see below).

The same principle applies for **modulators** where the time independent regression constants of modulator $u$ can be determined by some regression method [20][21]

$$(k_{u,1}, k_{u,2}, \dots, k_{u,h(u)}) = regression_u(x_{in1,u}(t_j), x_{in2,u}(t_j), x_{out,u}(t_j))\Big|_{t_j=t_{min}\dots t_{max}} \quad (27)$$

with a number of $h(u)$ regression constants $k_u$ of a modulator $u$.

The calibration of modulators can be achieved analogously to the one for bilateral couplings and a calibration set for a modulator then is defined by

$$x_{in1,u}(t_j)\ ;\ x_{in2,u}(t_j)\ ;\ x_{out,u}(t_j)\ \Big|_{t_j=t_{min}\dots t_{max}} \quad (28)$$

Like for bilateral couplings here all $x_{in1,u}(t_j)\ ;\ x_{in2,u}(t_j)$ can be calculated in a *forward pass* by (29) and all $x_{out,u}(t_j)$ can be calculated within a *backward pass* by (30) (see below).

The *net-function* introduced in following is based on the characteristics of NMPC-components connected with each other and thus constituting to a NMPC-network where an input state variable of a NMPC-component $u$ is simultaneously also the output state variable of another NMPC-component $v$. The output state variable of a NMPC-component $w$ is simultaneously the input state variable of a NMPC-component $v$.

Assuming that values for all regression constants $k_{u,1..h(u)}$ would be given and that $x_{in,u}$ in (26) and (28) is the output node of another NMPC-component $v$ then $x_{in,u}$ can be calculated by the net-function

$$x_{in,u}(t) = net_v(t) = \begin{cases} g_v(net_w(t), k_{v,1}, k_{v,2}, \dots, k_{v,h(v)}) & \text{with } x_{in,u} \text{ as output of a bilateral coupling } v \\ m_v(net_{w,1}(t), net_{w,2}(t), k_{v,1}, k_{v,2}, \dots, k_{v,h(v)}) & \text{with } x_{in,u} \text{ as output of a modulator } v \\ a_v(net_{w,1}(t), net_{w,2}(t), \dots, net_{w,j}(t)) & \text{with } x_{in,u} \text{ as output of a summator } v \\ p_u & \text{with } x_{in,u} \text{ as NMPC-parameter} \end{cases} \quad (29)$$

with generalized bilateral coupling function $g(t) \supseteq d(t)$, $g(t) \supseteq i(t)$, $g(t) \supseteq s(t)$.

For modulators, the net-function to calculate $x_{in,u}$ applies here for either inputs $x_{in1,u}$ and $x_{in2,u}$.

Further, assuming that $x_{out,u}$ in (26) and (28) is one input node of another NMPC-component $\bar{v}$ then $x_{out,u}$ can be calculated by the *net$^{-1}$-function* (inverse net-function)

$$x_{out,u}(t) = net_{\bar{v}}^{-1}(t) = \begin{cases} g_{\bar{v}}^{-1}(net_{\bar{w}}^{-1}(t), k_{\bar{v},1}, k_{\bar{v},2}, \dots, k_{\bar{v},h(\bar{v})}) & \text{with } x_{out,u} \text{ as input of a bilateral coupling } \bar{v} \\ x_{out,u}(m_{\bar{v}}^{-1}(net_{\bar{w}}^{-1}(t), k_{\bar{v},1}, k_{\bar{v},2}, \dots, k_{\bar{v},h(v)})) & \text{with } x_{out,u} \text{ as input of a modulator } \bar{v} \\ x_{out,u}(a_{\bar{v}}^{-1}(net_{\bar{w}}^{-1}(t))) & \text{with } x_{out,u} \text{ as input of a summator } \bar{v} \\ p_u & \text{with } x_{out,u} \text{ as NMPC-parameter} \end{cases} \quad (30)$$

whilst net$^{-1}$-function in all cases is assumed as calculable. This is non-trivial for the inverse functions $m_{\bar{v}}^{-1}$ and $a_{\bar{v}}^{-1}$ of the NMPC-operators, i.e. for NMPC-functions with more than one input state variable. A solution for this problem is shown in section IV.C.

For a NMPC-graph being defined as ***well formed*** all of following conditions need to be fulfilled:

- Each of its state variables needs to be computable by both (29) and (30).
- Each input state variable of any NMPC-component is connected to exactly either one output state variable or a NMPC-parameter.
- Each output state variable of any NMPC-component is connected to at least one input state variable or a NMPC-parameter.
- Each NMPC-function's output can be calculated directly or indirectly from at least one NMPC-parameter.

As the net-function and the inverse net-function allow to compute values for all input and all output state variables of any NMPC-function within a *well formed* NMPC-graph calibration is possible as soon as there is given a sufficient number of *NMPC-parameters* in column vector $P \subseteq X$ with $P(t(j))$, $j \in \mathrm{IN}$ and $j = 1...c$ where $c$ is a sufficient number of calibration records allowing to determine (26) and (28) assuming $c$ points in time in ascending order with $t_2(j_2) > t_1(j_1)$ given $j_2 > j_1$.

Since there may be connected more than one input state variable or NMPC-parameter to an output state variable of any NMPC-component $u$, the NMPC-graph may offer multiple output paths for choosing $net_{\bar{v}}^{-1}$ in (30). When implementing a software algorithm for calculation of an inverse net-function it is recommendable to randomly choose among multiple paths of one output node by selecting each of those paths with the same probability. This helps to achieve a well balanced consideration of all NMPC-components after many calibration cycles.

**Example:** Referring to section III.B there would apply following net-functions:

$$x_1(t) = p_1(t)$$
$$x_1(t) = net_1^{-1} = x_1(m_{4,2}^{-1}(t))$$
$$x_5(t) = net_5 = s_{5,2}(p_2(t))$$
$$x_5(t) = net_5^{-1} = x_5(a_6^{-1}(t))$$
$$x_6(t) = net_6 = s_{5,2}(p_2(t)) - m_{4,2}(p_1(t), p_3(t))$$
$$x_6(t) = net_6^{-1} = i_{3,6}^{-1}(p_3(t))$$

Note that calculating the inverse net-function of NMPC-operators like $m_{4,2}^{-1}(t)$ and $a_6^{-1}(t)$ is non-trivial and described in section IV.C.

*B. Proposing an algorithm for calibrating NMPC-functions*

The objective of this section is to show how all time independent regression constants $k_{u,1}, k_{u,2}, \dots, k_{u,h(u)}$ defined in (25) and (27) can be determined for all calibratable components in the NMPC-graph in order to fully define (19).

Following calibration algorithm is inspired by *back-propagation* algorithm for calibrating multilayered *artificial neural networks* [7]. However, instead of repetitively applying error-based weight adjustments it calibrates NMPC-functions by repetitively applying univariate (for bilateral couplings) and multivariate (for modulators) regression calculations where either none of those, some of those, or all of those calculations might be nonlinear.

Given a sufficient large number $c$ of calibration records following algorithm is proposed to find functions of (20) for $I$, $S$, $D$, and $M$ – excluding $A$ as it is not calibratable – in such way that *total calibration error* $\hat{\varepsilon}$ becomes minimized:

(1) Start.
(2) For all calibratable components $u$ in all functions $I$, $S$, $D$, and $M$ initialize regression parameters $k_{u,1}, k_{u,2}, \ldots, k_{u,h(u)}$ randomly while maintaining the constraints defined in section II.
(3) Provide NMPC-parameter calibration set with $c$ records containing values $p_u \in \mathrm{IR}$ for all $z \in IN_+$ NMPC-parameters
$$p_1(t_j)\,;\ p_2(t_j)\,;\ldots;\ p_z(t_j)\ \Big|_{t_j=t_{min}\ldots t_{max}\,;\ j=1\ldots c}$$
This is the calibration set the NMPC-model created by this algorithm is supposed to generalize.
(4) Count cycle.
(5) Calculate all regression constants $k_{u,v}$ for all calibratable NMPC-functions due to (25) and (27) and replace old regression constants by the new ones. For each regression function calculate values of all its input variables by forward pass via (29) and all values of its output variable by backward path via (30).
(6) For each NMPC-function $u$ enforce the function's constraints defined in section II.
(7) For each NMPC-parameter $p_u$ of all $z$ NMPC-parameters: calculate error over all $c$ calibration records by applying the net-function in (29):
$$\varepsilon_u = maximum(\left|p_u(t_j) - net_u(t_j)\right|\Big|_{t_j=t_{min}\ldots t_{max}\,;\ j=1\ldots c})$$
(8) Determine total calibration error
$$\hat{\varepsilon} = maximum(\varepsilon_1, \varepsilon_2, \ldots, \varepsilon_z)$$
over all of the $z$ NMPC-parameters within the NMPC-graph.
(9) If number of cycles computed in (4) is beyond acceptable limit or $\hat{\varepsilon}$ is sufficiently low, leave loop via step (14).
(10) Modify *simulated annealing* temperature [22].
(11) Apply *simulated annealing algorithm* [22]: Memorize current values of all regression parameters $k_{u,v}$ of all calibratable NMPC-functions, then modify the values of all of them, randomly. Obtain new $\hat{\varepsilon}'$ by re-computing (7) and (8). If $\hat{\varepsilon}'$ does not fulfill the *simulated annealing* step acceptance criterion reject the random modification of all $k_{u,v}$ and restore the memorized values, otherwise accept the random modification.
(12) For each NMPC-function $u$ enforce the function's constraints listed in section II.
(13) Go to step (4).
(14) Conserve all $k_{u,v}$ for solving state equation (19). Thus, the state equation becomes fully defined and will be called *NMPC-model*.
(15) End.

The convergence of reducing $\hat{\varepsilon}$ to minimum by this algorithm is not proven within this article.

*C. Calculating inverse functions of functions with multiple input variables*

The inverse function of a NMPC-function with exactly one input node $g_{\bar{v}}^{-1}(net_{\bar{w}}^{-1}(t))$ in (30) like *integrative*, *synchronous*, and *differential couplings* can be easily determined, because $g_{\bar{v}}$ and thus also $g_{\bar{v}}^{-1}$ are strictly monotone functions yielding one output value for one input value over their whole definition interval.

On the contrary, in (30) the inverse function of a NMPC-function with multiple input variables like *summators* with $a_{\bar{v}}^{-1}(net_{\bar{w}}^{-1}(t))$ or *modulators* with $m_{\bar{v}}^{-1}(net_{\bar{w}}^{-1}(t))$ in spite of being strictly monotone are not real functions in a mathematical sense as they are yielding multiple valid input variable value sets for exactly one output value. The basic idea now is to use input values previously calculated in a forward calculation pass for dissolving this ambiguity for both modulator and summator.

When applying the inverse function during a backward calculation pass in an ideally calibrated NMPC-model its results at the output variable of any NMPC-function would not deviate from the same value calculated by a forward pass.

However, if it does as being the case in a non-ideal calibration state the inverse function can be calculated in such way that it approximates most closely the result of the last forward pass.

The two following sections describe why the inverse function results in an ambiguity for operators and how it can be dissolved for modulators and summators in a reasonable way.

*3) Calculating inverse function of a NMPC-modulator*

The surface of the strictly monotone function of a modulator can be expressed by position vector $\vec{r}$ with

$$\mathrm{IR}^3: \qquad \vec{r} = \begin{pmatrix} x_{in,1} \\ x_{in,2} \\ m(x_{in,1}\ ,\ x_{in,2}) \end{pmatrix} \tag{31}$$

with the characteristics described in section II.C,2). Determining the modulator's input variable values by inverse function $\begin{pmatrix} x_{in,1} \\ x_{in,2} \end{pmatrix} = m^{-1}(x_{out})$ geometrically means to find the intersection between $\vec{r}$ and a plane $\begin{pmatrix} x_{in,1} \\ x_{in,2} \\ x_{out} = constant \end{pmatrix}$ for a given $x_{out}$ which for a strictly monotone plane in either of the input variables $x_{in,1}$ and $x_{in,2}$ results in a geometric line

$$\mathrm{IR}^2: \quad \vec{l}_{in} = \begin{pmatrix} x_{line_in,1} \\ x_{line_in,2} \end{pmatrix} = \begin{pmatrix} x_{line_in,1} \\ m^{-1}(x_{out}, x_{line_in,1}) \end{pmatrix} = \begin{pmatrix} m^{-1}(x_{out}, x_{line_in,2}) \\ x_{line_in,2} \end{pmatrix} \tag{32}$$

constituted by an infinite number of points. This solution is undesired, because for calculating a modulator's inverse function $m_{\bar{v}}^{-1}(net_{\bar{w}}^{-1}(t), k_{\bar{v},1}, k_{\bar{v},2}, \ldots, k_{\bar{v},h(v)})$ in (30) a modulator must render exactly one point $\begin{pmatrix} x_{point_in,1} \\ x_{point_in,2} \end{pmatrix}$ instead of a line $\vec{l}_{in}$.

However, the ambiguity of $\vec{l}_{in}$ can be dissolved by taking into account additional information as resulting from the algorithm proposed in section IV.B:

During step (7) all functions

$$net_u(t_j)\Big|_{t_j = t_{min}..t_{max};\, j=1..c}$$

are calculated based on all of their input variables for all given points in time. When doing that for modulator $\bar{v}$ it is useful to memorize for each $t_j$ a point

$$\mathrm{IR}^2: \quad \vec{q}_{\bar{v}}(t_j) = \begin{pmatrix} x_{mem_in,1,\bar{v}}(t_j) \\ x_{mem_in,2,\bar{v}}(t_j) \end{pmatrix} \tag{33}$$

representing the modulator $\bar{v}$ 's input variable values.

At algorithm step (5) which implies calculation of inverse functions due to (30), for modulators such memorized point $\vec{q}_{\bar{v}}(t_j)$ from the previous cycle can be used to determine unambiguous input variable values by selecting the one point of $\vec{l}_{in}(t_j)$ being closest to $\vec{q}_{\bar{v}}(t_j)$ by calculating

$$minimum\left( \left| \vec{l}_{in}(t_j) - \vec{q}_{\bar{v}}(t_j) \right| \right) \rightarrow \begin{pmatrix} x_{point_in,1,\bar{v}}(t_j) \\ x_{point_in,2,\bar{v}}(t_j) \end{pmatrix} \tag{34}$$

where $x_{point_in,1,\bar{v}}(t_j)$ and $x_{point_in,2,\bar{v}}(t_j)$ represent the input values rendered by modulator $\bar{v}$ 's inverse function where each of them corresponds to $net_{\bar{v}}^{-1}(t)$ in (30).

#### 4) Calculating inverse function of a NMPC summator

Like for modulators also for summators all of them are calculated during step (7)

$$net_u(t_j)\Big|_{t_j = t_{min}..t_{max};\, j=1..c}$$

based on all of their input variables for all given points in time in course of a forward pass. When doing that for the summator $u$ it is useful to memorize for each $t_j$ the summators output value $x_{mem_out,u}(t_j)$ and for each of its input variable values $x_{mem_in1..y,u}(t_j)$ with $y$ input variables, $y \in \mathrm{IN}_+$.

When calculating the backward pass leading to a summator $u$ as implied in step (5) and resulting in $x_{out,u}(t_j)$ the values memorized from the previous cycle in step (7) can be used to calculate the deviation between previous cycle and current cycle by

$$\Delta x_{out,u}(t_j) = x_{out,u}(t_j) - x_{mem_out,u}(t_j) \ . \tag{35}$$

This deviation now can be propagated to the current input variables $x_{in1..y,u}(t_j)$ by distributing the output deviation equally over the input variables by

$$x_{in1..y,u}(t_j) = x_{mem_in1..y,u}(t_j) + \frac{\Delta x_{out,u}(t_j)}{y} \ . \tag{36}$$

If during many calibration cycles the value for $x_{out,u}(t_j)$ as calculated during forward pass approximates each time closer the one calculated during backward pass the deviation $\Delta x_{out,u}(t_j)$ becomes smaller and smaller. This way also the input variable values $x_{in1..y,u}(t_j)$ calculated during forward pass are approaching that ones calculated during backward pass.

## V. Calibration as Hypothesis Testing

A specific NMPC-graph can be understood as hypothesis about the causal dynamic structure of some system producing the calibration data.

Each of $c$ calibration records is supposed to match state equation (20) with given values for $P \subseteq X$. During calibration each regression constant $k_{u,v}$ is one free variable within the state equation to be calculated in the course of solving the inverse problem as shown in section 0. For not being underdetermined there need to be at least as many equations populated by values for $P$ as free variables, i.e. at least as many calibration records as regression constants. A good base for calibration will be given if $c$

- is much greater than the amount of regression constants $k_{u,v}$ .
- is sufficiently large for covering all relevant system behavior.

If the NMPC-graph is representing the system's nature well calibration error $\hat{\varepsilon}$ of its corresponding NMPC-model is expected to result small after termination of the algorithm described in section IV.B. It expresses the maximum absolute deviation of any of the given NMPC-parameter values versus the ones calculated based on (20) over all calibration records. If it results small, i.e. as small portion of all of the NMPC-parameter value ranges $[P_{min,u}, P_{max,u}]$ the calibration succeeded and the NMPC-model is likely to represent the system well. In such case the hypothesis given by a NMPC-graph can be assumed as relevant.

## VI. Solving the State Equation

Once, regression parameters $k_{u,1}, k_{u,2}, \dots, k_{u,h(u)}$ of all calibratable NMPC-functions $u$ as calculated in (25) and (27) have been determined by calibration, the NMPC-parameter values within the nonlinear state equation as introduced in section III.A can be solved numerically by a fixed-point iteration process. However, as net-functions of the NMPC-model generally are multivariate and potentially nonlinear a satisfying fixed-point iteration algorithm will be non-trivial.

There are numerous possible fixed-point iteration methods to solve the nonlinear state equation [23]. In following there will not be presented a fully specified algorithm but the properties it needs to have.

One possible starting method might be a Gauss-Seidel process [25][26] which seeks to approximate the solution value for $p_v(t)$ of the current NMPC-model calibration cycle within an interval $[P_{v_min}, P_{v_max}]$ for each of all $z$ NMPC-parameters by

$$\begin{aligned}&{}_{(w+1)}p_v(t) = \\ &\quad net_v({}_{(w+1)}p_1(t), {}_{(w+1)}p_2(t), \ldots, \\ &\qquad {}_{(w)}p_j(t), \ldots, {}_{(w)}p_{z-1}(t), {}_{(w)}p_z(t))\big|_{v=1\ldots z;1\leq j\leq z}\end{aligned} \tag{37}$$

where $w+1$ is current and $w$ is previous iteration, $v = 1 \ldots z$ identifies the currently calculated NMPC-parameter, and $net_v$ is the net-function determining the value of the current NMPC-parameter depending on all other NMPC-parameters and, in recursive systems, also from itself. The starting value ${}_{(0)}p_v(t)$ is $p'_v(t)$ which is the value determined during the previous NMPC-model calibration cycle or some guess value for the first calibration cycle.

Convergence of the iteration process is given as long as

$$\left|{}_{(w+2)}p_v(t) - {}_{(w+1)}p_v(t)\right| < \left|{}_{(w+1)}p_v(t) - {}_{(w)}p_v(t)\right| \tag{38}$$

for all $w$ iterations. Iteration should be aborted when

$$\left|{}_{(w+1)}p_v(t) - {}_{(w)}p_v(t)\right| < e \tag{39}$$

with $e$ as acceptable iteration error.

As soon as (38) is not fulfilled anymore, an alternative iteration method needs to be applied for finding a new ${}_{(w+1)}p_v(t)$. In such case (37) will be only used to test abortion criterion by (39). Starting with the value for $p'_v(t)$ such method will probe different values for ${}_{(w+1)}p_v(t)$ within $[P_{v_min}, P_{v_max}]$ trying to either fulfill (39) or $\left|{}_{(w+1)}p_v(t) - {}_{(w)}p_v(t)\right| \rightarrow minimum$ with the closest distance to the value of $p'_v(t)$ possible. Such method is expected to be computationally costly in comparison to a converging fixed-point iteration method, but is not limited by convergence criteria.

## VII. Outlook

The algorithm described so far is suited to serve as implementation base for a prediction software solution. Such will consist of two parts:

The first part will be a calibration engine which consumes a formal description of a NMPC-graph and historic time-series data measured from the scenario which later on shall be predicted. From that input the software will derive the NMPC-model by a calibration process as described in section 0.

The second part of such software solution will be a time-series predictor. It will consume a previously calibrated NMPC-model and time-series scenario data for predicting the scenario's behavior in future. From that it derives the time-series prediction due to the principle described in section VI.

Such a software solution already has been implemented and is available as freeware edition [27]. A user manual containing a complete description for calibrating and predicting a scenario according to the one presented in III.B also has been published [28].

## Appendix – Glossary

***Terms of the NMPC methodology***

| | |
|---|---|
| backward pass | Calculating the output of a *NMPC-component* from backward by solely applying *inverse net-functions* throughout the NMPC-network. |
| bilateral coupling | Superclass of following couplings: *differential*, *integrative*, and *synchronous coupling*. |
| calibration | Derivation of a *NMPC-model* by (a) a *NMPC-graph* and (b) given calibration records via a complex regression algorithm, potentially under the condition of multivariate, nonlinear, temporal, and recursive network structure.<br><br>The characteristics of this method show several similarities to *backpropagation* algorithm for *artificial neural networks*, e.g. its iterative nature and the challenge to overcome *local error minima* in order to approach the *global error minimum*. |
| differential coupling | *NMPC-component* where the growth of one output *state variable* depends potentially nonlinearly on the acceleration of one input *state variable* over the *independent variable*. |
| dynamic | Time related behavior, see also *temporal*. |
| forward pass | Calculating the input of a *NMPC-component* from forward by solely applying *net-functions* throughout the NMPC-network. |
| independent variable | A *NMPC-graph* describes a dynamic system whose behavior depends on one *independent variable*, i.e. this variable is independent from any other variable or constant within the *NMPC-model*. It is also the *independent variable* used in *bilateral couplings*. Typically, this *independent variable* represents the time measure. |
| integrative coupling | *NMPC-component* where the growth of one output *state variable* depends potentially nonlinearly on the level of one input *state variable* over the *independent variable*. |
| inverse net-function | Abstract function which calculates some state variable value within the NMPC-network by applying a cascade of inverse *NMPC-functions* starting with one or more *NMPC-parameters* as inputs. |
| latent variable | All *state variables* whose corresponding value cannot or is not being measured within their corresponding real world system. The meaning is the same like for the conventions of *Structural Equation Modeling*. |
| m-function | Monotone function at the base of each bilateral coupling type. |
| manifest variable | All *state variables* whose corresponding value is being measured within their corresponding real world system. The meaning is the same like for the conventions of *Structural Equation Modeling*. |
| modulator | *NMPC-component* where the value of one output *state variable* depends potentially nonlinearly on the values of two input *state variables*. |
| monotone function | In the context of *NMPC-graph* a mathematical function where a rising input variable results in a rising output variable at every point of the definition interval of the function. |
| net-function | Abstract function which calculates some *state variable* value within the NMPC-network by applying a cascade of *NMPC-functions* starting with one or more *NMPC-parameters* as inputs. |
| NMPC-component | One of following components within any *NMPC-graph*: Either a *NMPC-parameter*, an *operator*, or a *bilateral coupling*. |
| NMPC-function | Either an *operator* or a *bilateral coupling*. |
| NMPC-graph | Path diagram of a **N**onlinear **M**onotone **P**arameter **C**oupling system description, exclusively formed by *NMPC-components* and their mutual connections. |
| NMPC-model | A strictly mathematical representation of the *NMPC-graph* where all constants of all equations, i.e. the shape of all of its mathematical functions, are fixed and fully specified. *Independent* and *state variables* remain fluctuating.<br><br>As soon as a *NMPC-model* contains *integrative* and/or *differential couplings* it forms an ordinary differential equation (ODE) system over the *independent variable*. |
| operator | One of following components within a *NMPC-graph*: either a *summator* or a *modulator*. |
| (NMPC-)parameter | *NMPC-component* representing a *state variable*. (NMPC-)*parameters* represent some or all of the *manifest* or *latent* state variables. All *manifest variables* need to be presented to the NMPC-model by parameters.<br><br>Typically, *parameters* are placed within *NMPC-graph* in such way that they represent observables or other meaningful values. |
| relevant hypothesis | A *NMPC-graph* is termed to be a *relevant hypothesis* as soon as its best calibrated *NMPC-model* can reproduce the *NMPC-parameter* values of all calibration records sufficiently well. |
| state variable | All variables which are input to or result from the equations within the *NMPC-model*. Each s*tate variable* holds a real number value describing the state of the dynamic system dependent on the *independent variable* and *parameters*, i.e. other *state variables*. The sets of *latent* and *manifest variables* are exclusively composing the set of *state variables*. |
| summator | *NMPC-component* where the value of one output *state variable* is the linear sum of two up to many values of its input *state variables*. |
| synchronous coupling | *NMPC-component* where the growth of one output *state variable* depends potentially nonlinearly on the growth of one input *state variable* over the *independent variable*. |
| temporal | Time related, typically integrative or differential functions. |

***Mathematical placeholders***

| | |
|---|---|
| $a_u(...)$ | *Monotone function* of a *summator u*, always defined as linear sum of its signed input *state variable* values. |
| $a^{-1}$ | Inverse function of a *summator*. |
| $\tilde{a}_{u,v}(x)$ | Signed input *v* of *summator u* determining whether input is added positively or negatively. |
| $c$ | Number of *calibration* records. |
| $d_u(x)$ | *Differential coupling* function of component *u*. |
| $\tilde{d}(x)$ | *Monotone function* of a *differential coupling*, potentially with nonlinear shape. |
| $e$ | Iteration abortion limit for the difference between values of two successive fixed-point iteration steps or Gauss-Seidel process steps of one *NMPC-parameter*. |
| $\varepsilon_u$ | Calibration error over all calibration records for *NMPC-parameter u* during one *calibration* cycle. |
| $\hat{\varepsilon}$ | Calibration error over all calibration records for all *NMPC-parameters* during one *calibration* cycle. |
| $\hat{\varepsilon}'$ | Calibration error over all calibration records for all *NMPC-parameters* after simulated annealing random modifications during one *calibration* cycle. |
| $g_u$ | Generic monotone function generalizing *NMPC-function* for a *bilateral coupling* $i_u(x)$, $d_u(x)$, $s_u(x)$ in one input *state variable* and with one output *state variable* whose shape is fully defined by its regression constants $k_{u,v}$. |
| $h(u)$ | Number of regression parameters for a calibratable *NMPC-function* of *NMPC-component u.* |
| $i_u(x)$ | *Integrative coupling* function of component *u*. |
| $\tilde{i}(x)$ | *Monotone function* of an *integrative coupling*, potentially with nonlinear shape. |
| $j$ | Index of some variable or function, $j \in \mathrm{IN}$ |
| $k_{u,v}$ | Regression constants of either a generic *monotone function* $g_u$ for a *bilateral coupling* or a *modulator* $m_u$. |
| $l$ | Index of some variable or function, $l \in \mathrm{IN}$ |
| $\vec{l}_{in}$ | An intersection line on a *modulator's* plane which is obtained during calculation of the inverse function of a *modulator*. |
| $m(x)$ | General *monotone function*, potentially with nonlinear shape. |
| $m(x_{in1}, x_{in2})$ | Function of a *modulator* being monotone and potentially nonlinear in each of its input variables. |
| $m^{-1}$ | Inverse function of a *modulator*. |
| $n$ | Number of *state variables* within the *NMPC-graph*. |
| $p_u$ | *NMPC-parameter u* within the *NMPC-graph*. |
| $P$ | Column vector of all *NMPC-parameters* $p_v$ within the *NMPC-graph*. |
| $q$ | Maximum number of choices to combine two *state variables* as inputs for one *modulator*. |
| $\vec{q}_{\bar{v}}$ | Memorized input *state variable* value set for supporting subsequent calculation of a *modulator's* inverse function. |
| $r$ | A row index within the matrices corresponding to *state variable* $x_r$. |
| $\vec{r}$ | Position vector describing the plane of a *modulator*. |
| $s_u(x)$ | *Synchronous coupling* function of component *u*. |
| $\tilde{s}(x)$ | *Monotone function* of a *synchronous coupling*, potentially with nonlinear shape. |
| $t$ | *Independent variable*, typically representing time. |
| $u$ | Index uniquely identifying a *NMPC-component* within an NMPC-graph, $u \in \mathbb{N}_+$. |
| $v$ | Index of some variable or function, $v \in \mathbb{N}_+$. |
| $w$ | Index of some variable or function, $w \in \mathbb{N}_+$. |
| $x_{in}$ | Input *state variable* used in the definition of a *NMPC-component*. |
| $x_{out}$ | Output *state variable* used in the definition of a *NMPC-component*. |
| $x_v$ | *Latent* or *manifest state variable* within the *NMPC-graph*. |
| $X$ | Column vector of all *state variables* $x_v$ within the *NMPC-graph*. |
| $y(u)$ | Number of input *state variables* $x_v$ of *NMPC-function u* with $y \in \mathbb{N}_+$. |
| $z$ | Number of *NMPC-parameters*. |